\documentclass[a4paper,aps, prl,amsmath,amssymb, superscriptaddress, showpacs, twocolumn, floatfix]{revtex4}
\usepackage{graphicx, braket, mathrsfs, nicefrac, epsfig, multirow, ulem, verbatim, subfigure}

\begin{document}
\title{Topological properties of superconducting junctions}
\author{D. I. Pikulin}

\affiliation{Instituut-Lorentz, Universiteit Leiden, P.O. Box 9506, 2300 RA Leiden, The Netherlands}

\author{Yuli V. Nazarov}

\affiliation{Kavli Institute of Nanoscience, Delft University of Technology, Lorentzweg 1,
2628 CJ Delft, The Netherlands}

\begin{abstract}
Motivated by recent developments in the field of one-dimensional topological superconductors, we investigate the topological properties of s-matrix of generic superconducting junctions where dimension should not play any role. We argue that for a finite junction the s-matrix is always topologically trivial. We resolve an apparent contradiction with the previous results by taking into account the low-energy resonant poles of s-matrix. Thus no common topological transition occur in a finite junction. We reveal a transition of a different kind that concerns the configuration of the resonant poles.
\end{abstract}
\pacs{71.10.Pm, 74.45.+c, 03.67.Lx, 74.90.+n}
\maketitle

Superconducting junctions, including superconducting-normal (SN) ones where dissipative conduction can take place and superconducting-superconducting (SS) ones where a discrete spectrum of bound Andreev states is formed, have been in focus of condensed-matter research for almost fifty years \cite{Josephson,
Nazarov1}. An indispensable compact approach to superconducting junctions employs a scattering matrix that relates incoming and outgoing wave amplitudes that obey the Bogolyubov-deGennes (BdG) equation \cite{Shelankov,Beenakker2}. The beauty and power of this approach stems from its ability to incorporate numerous microscopic details in a compact form of the scattering amplitudes. Straightforward extensions  permit to include magnetism, spin-orbit interaction, non-trivial superconducting pairing \cite{de Jong}. The s-matrix approach can be easily combined with semiclassical treatment of electron transport in the framework of a quantum circuit theory \cite{Nazarov1}.

Recent developments in the field of superconductivity require revision of the common assumptions concerning the structure and properties of the scattering matrix of a superconducting junction. Kitaev in 2000 has suggested a model 1d p-wave superconductor \cite{Kitaev} that exhibits a topological order. It has been shown recently that the same topological order can be realized in more realistic systems that combine spin magnetic field \cite{Sau} with  strong spin-orbit interaction \cite{Lutchyn,Oreg1}. Similar situation would occur in a superconductor on the top of topological insulator or half-metal \cite{Duckheim}. The relevance of these developments for generic superconducting junctions is not immediately relevant. Indeed, the general properties of those are not supposed to depend on dimension \cite{Beenakker3}, while topological ordering considered is specific for one dimension \cite{Ryu} thus suggesting that the topological properties are not at all manifested in junctions. However, a number of spectacular predictions and device schemes that relate the topology  and junction properties has appeared in the last years. Those include: prediction of so-called $4\pi$ periodic Josephson effect \cite{Fu2, Lutchyn, Oreg},  formulation of a criterion for topological transition in terms of reflection matrix of a junction \cite{Merz}, proposals of  topological qubits based on majorana bound states \cite{Kitaev,Oreg} as well as their readout with qubits  of different type \cite{Hassler}.

Thus motivated, we have performed a topological analysis of a general BdG scattering matrix concentrating on energy dependence of its eigenvalues. This rather elementary analysis shows that $i$. there are  topologically non-trivial (TNT) s-matrices characterized by real eigenvalues at zero energy, $ii$. there are topologically non-trivial trajectories (TNTT) in the space of topologically trivial (TT) s-matrices, that pass a matrix with real eigenvalues at $E=0$  odd number of times.

TNT would correspond to a "topological" SN junction \cite{Beenakker4}, while TNTT would explain $4\pi$-periodicity of Josephson effect in SS junctions \cite{Lutchyn, Fu2}. Albeit the same topological reasoning implies topological triviality of  all physical s-matrices: {\it there are no TNT neither TNTT}. This forms a paradox that is resolved by recognizing a potentially sharp energy dependence of a s-matrix near zero energy. Such energy dependence is due to resonant poles \cite{Wigner} that manifest formation and coupling of zero-energy quasilocalized states. With this, we reconcile the predictions of \cite{Lutchyn, Fu2}, show the absence of a common topological transition and reveal topological transitions related to the resonant poles.

\begin{figure}
\centerline{\includegraphics[width=0.9\linewidth]{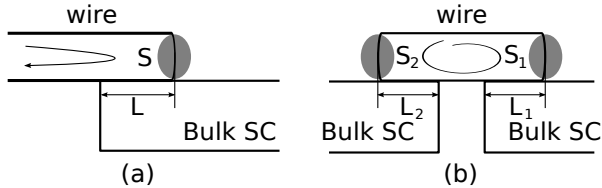}}
\caption{Setups to illustrate general topological properties of BdG s-matrices. a. Finite-length wire with strong spin-orbit coupling on the top of superconducting lead forming a SN junction. b. Finite-length wire  between two  superconductors forming SS junction.
Grey ellipses indicate "buried" zero-energy states.
}
\label{fig_setup}
\end{figure}

We illustrate these results with two minimal setups, SN and SS junctions (Fig. \ref{fig_setup}), where a single-channel wire with strong spin-orbit coupling and subject to magnetic field  is brought in contact with a bulk superconductor. The Hamiltonian description of this situation is found in \cite{Lutchyn}. In distinction from \cite{Lutchyn}, we assume finite length of the contact. The solutions of BdG equation for a single channel encompass spin and electron-hole degree of freedom so that the minimal single-channel scattering matrix is $4\times4$.
The parameter space of the model that includes the superconducting gap, chemical potential, strength of spin-orbit interaction, and magnetic field, can be separated into two ranges: "topological" and non-"topological".

Let us consider a general s-matrix of a SN junction assuming no symmetries. The only constraint on
such matrix stems from the structure of BdG equation: its Hamiltonian satisfies $\hat{H}^{*} = -
\tau_1 \hat{H} \tau_1$, where the operator $\tau_1$ switches electrons and holes. The constraint is
convenient to represent in so-called Majorana basis \cite{Majorana} where the Hamiltonian is antisymmetric and the
scattering matrix satisfies $S(E)=S^*(-E)$ , $E$ being energy counted from the
chemical potential of the superconductor. We will consider only energies $E$  within the
 energy gap of the bulk superconductor. In this case, there are no scattering waves in the bulk of
superconductor, the matrix $\hat{S}$ is in the basis of normal-metal scattering waves
satisfying unitary condition.

\begin{figure}
\centerline{\includegraphics[width=0.9\linewidth]{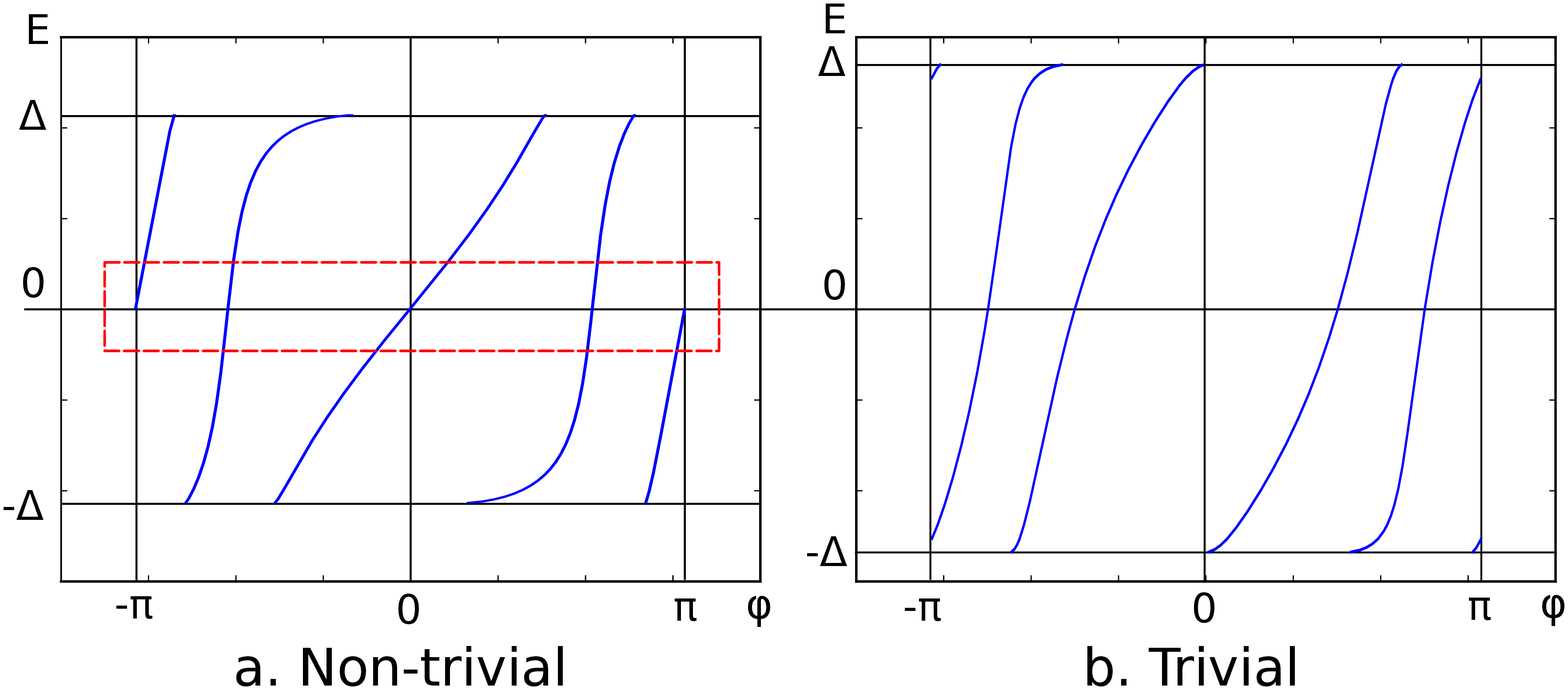}} \caption{ Energy dependence of
s-matrix eigenvalues. (a) Topologically non-trivial (TNT) case, corresponding to the "topological" parameter range  in \cite{Lutchyn}. (b) Generic topologically trivial (TT) case. (Numerical results for the setup in Fig. 1a in the limit $L\to \infty$.)}
\label{fig_SM}
\end{figure}

Let us concentrate on (continuous) energy dependence of the matrix eigenvalues $e^{i\chi(E)}$. That can be represented as
a manifold of curves in $\chi-E$ plane (Fig. \ref{fig_SM}). The BdG constraint implies that if
a point $(\chi,E)$ belongs to the manifold, the inverted point $(-\chi,-E)$ belongs to it as well.
These two points can belong to either the same curve or to two distinct curves. In the first case, the
curve is topologically distinct: it is forced to pass either $\chi=0$ or $\chi=\pm \pi$ at zero
energy. If two such curves pass the same point, they can be deformed by continuous change of
Hamiltonian parameters into a pair of trivial curves. However, a single curve is topologically
stable: the fact it passes the point cannot be changed by Hamiltonian variations.
We note that the dimension of the physical s-matrices can be always chosen even.
With all this, all s-matrices can be separated onto two classes. Topologically trivial (TT) matrices have no topologically distinct curves while topologically non-trivial (TNT) have two topologically
distinct curves passing respectively $\chi=0$ and $\chi=\pm \pi$ at $E=0$. Indeed, at zero energy s-matrices are real forming O(2N) group. TT matrices  belong to SO(2N) subgroup of O(2N), while TNT belong to O(2N)/SO(2N). The matrices from these distinct submanifolds cannot be continuously deformed into one another: indeed, at $E=0$ $\det(TT)=1$ while $\det(TNT)=-1$.

\begin{figure}
\centerline{\includegraphics[width=0.9\linewidth]{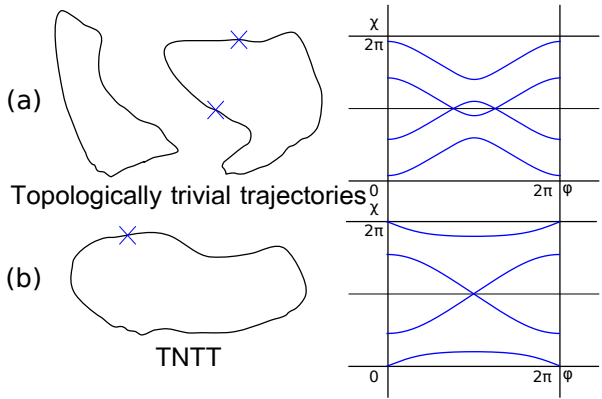}} \caption{Topological classes of trajectories in the space of TT s-matrices. A trajectory is topologically non-trivial (TNTT) provided it passes  the matrix with two degenerate real eigenvalues {\it odd} number of times. Illustration: the dependencies of 
eigenvalues of the scattering matrix characterizing the SS junction on superconducting phase difference
$\phi$ for (a) non-"topological" and (b) "topological" parameter ranges.} \label{fig4}
\end{figure}

This classifies s-matrices of SN junction. An SS junction is characterized by a combination of two s-matrices (Fig. \ref{fig4}). The spectrum of Andreev states of the junction as function of superconducting phase difference $\phi$ is obtained from the equation \cite{Beenakker2}
\begin{equation}
\label{eq:spectrum}
0={\rm det}\left(\hat{1} - \hat{S}\right); \; \hat{S} = \hat{s}_1 e^{i \phi\tau_3/2} \hat{s}_2 e^{-i\phi\tau_3/2},
\end{equation}
$\tau_3$ being Nambu matrix distinguishing electrons and holes. It is instructive to note that the unitary matrix $\hat{S}(\phi)$ satisfies the same BdG constraint as an $SN$ s-matrix. Therefore, the above topological classification applies to SS junctions as well.

\begin{figure}
\centerline{\includegraphics[width=0.9\linewidth]{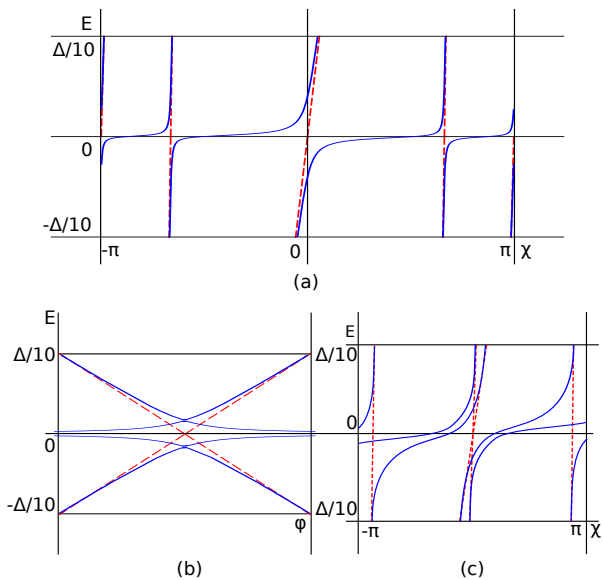}}
\caption{(a): Energy dependence of  eigenvalues for NS junction 
in a narrow energy interval illustrates the topological triviality of $s$-matrix  
for finite length of the contact (L=$7$ in units of \cite{Lutchyn}). Dashed lines: "high"-energy TNT eigenvalues. We see the reconnection of neighboring eigenvalues.
(b):  Andreev levels in SS junction versus superconducting phase difference at $L_1 =L_2=7$ (solid lines) as compared to TNTT case at $L{1,2}=\infty$ (dashed lines). (c) Energy dependence of eigenvalues for case (b) and $\phi=\pi$. Dashed lines:  TNTT case.}
\label{fig3}
\end{figure}

In this respect it is crucial to note another topological property that concerns continuous
one-parameter closed manifolds of TT matrices (trajectories). Intuitively,
eigenvalues of a generic matrix "repel" each other and never come together. This applies  
to BdG matrices expect a special situation: $E=0$ and  real eigenvalues. Owing to this peculiarity, a trajectory in matrix space can in
principle pass a matrix where two eigenvalues, say, $+1$, are the same. It turns out that the
trajectories of the kind can be separated onto two topological classes that differ by parity of the
number of passes. (Fig. \ref{fig3}) To see the possibility for odd number of passes, let us take a closed trajectory with a single pass
and concentrate on two eigenvectors corresponding to the eigenvalue $+1$. In this situation, if the parameter cycles over the trajectory, a  given eigenvector is transformed not to itself but rather to its orthogonal counterpart, this guarantees the stability of this topologically non-trivial
trajectory (TNTT).

Let us understand the results of\cite{Lutchyn,Oreg,Fu2} in terms of the above classification. Without going into details, we enunciate that TNT s-matrices are realized in the "topological"  parameter range. The TNTT give the topological explanation of the $4\pi$  Josephson effect described in these articles. The trajectory parameter in this case is the superconducting phase difference $\phi$.


However, similar topological considerations show that {\it no} physical s-matrix belongs to TNT class, {\it neither} any closed trajectory in parameter space is a TNTT. To prove this, let us start with a common (finite) junction manifesting no exotic properties. For our examples, this may correspond to zero magnetic field and zero spin-orbit interaction. The s-matrix at this parameter choice as well as all trajectories are topologically trivial. Since there is no continuous way to tune scattering matrix from TT to TNT class,
the s-matrix will stay trivial at any strength of magnetic field/spin-orbit interaction, even in the "topological" parameter range.  This proof is in apparent contradiction with the predictions mentioned \cite{Lutchyn,Fu2}, this forming a \textit{paradox} that motivated us for the present research.

Prior to presenting the solution of the paradox, let us mention that the absence of TNTT resolves an annoying problem that concerns the parity of particle number of the ground state of the SS junction. The level crossings
at $E=0$ are known in the context of ferromagnetic SS junctions. Upon passing the crossing, it becomes energetically favourable to put a single polarized quasiparticle to the junction \cite{Falko}. Therefore, the parity of the ground state must be different at two sides of the crossing.
Odd number of crossings at a closed curve implies indefinite parity of the ground state: a situation that is annoyingly difficult to comprehend.

To see how the paradox is resolved, let us consider numerical results for a finite SN junction
in "topological" parameter range.(Fig. 4a) If the results are plotted at energy scale of the superconducting gap,
the pattern of energy dependent eigenvalues is apparently of TNT type as in Fig. 2a. However, replotting the results near $E=0$ at smaller scale reveals topological triviality (cf. Fig. 4a and Fig. 2b). The eigenvalues move fast in the vicinity of $E=0$ reconnecting the branches visible at larger energy scale in a rather unexpected way.
The typical energy scale of such reconnection is small depending exponentially on the contact length $L$, and shrinks to zero at $L \to \infty$. This solves the paradox.

The adequate description of the situation combines a smooth energy dependence of s-matrix at $E\simeq \Delta$ with a
pole or poles that are anomalously close to $E=0$. Let us consider a single pole. The BdG
constraints restrict it to purely imaginary energy, $-i\Gamma \ll \Delta$.  The s-matrix reads
\begin{equation}
\hat{s}=\left(\hat 1 - \left(\frac{\epsilon-i\Gamma}{\epsilon+i\Gamma}
-1\right)|\Psi><\Psi|\right)\hat{S}_0, 
\end{equation}
where $\Psi$ is the eigenvector associated with the resonant level and $\hat S_0$ is the matrix,
with smooth  energy dependence to disregard at $E \simeq \Gamma$. The eigenvalues in this energy range are determined from equation $ \epsilon/\Gamma=\sum_k |\Psi_k|^2 \cot (\chi_k -\chi^{(0)}_k ),$ $\exp(i\chi^{(0)}_k)$ being "high-energy" ($|E| \gg \Gamma$) eigenvalues of $S_0$ . They follow the pattern
in Fig. \ref{fig3} connecting neighboring "high-energy" eigenvalues, $\exp(i\chi^{(0)}_k) \to
\exp(i\chi^{(0)}_{k+1})$. This guarantees that the  total shift of  phases of all eigenvalues upon
crossing a single pole equals $2\pi$.  Physically, the pole is associated with a quasi-localized
zero-energy state being formed at the far end of the wire. If the contact length exceeds the
localization length, this state is efficiently "buried" ($\Gamma \ll \Delta$) in the superconductor
and hardly accessible for incoming electron or hole waves  except $E=0$ when the scattering of the waves become resonant.  Andreev conductance of the junction is expressed as $G_A = G_Q {\rm Tr} \left(\tau_3\hat{s} \tau_3\hat{s}^\dagger\right)$. In the resonant energy interval, the energy
dependence of the conductance assumes a universal form $G_A(E) = G_A + \frac{\Gamma^2}{E^2 +\Gamma^2} (G_A(0) -G_A),$ $G_A(0),G_A$ being its values at $E=0,|E| \gg \Gamma$ that depend on details of the junction.

\begin{figure}
\centerline{\includegraphics[width=0.9\linewidth]{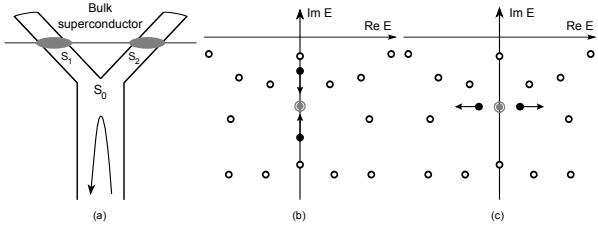}} \caption{(a): Fork SN junction to illustrate topological transitions concerning the resonant poles.  (b,c): Configurations of the resonant poles in the complex energy plane
(b) before and (c) after a transition. At the transition point, the poles are degenerate (double gray circle)} \label{fig5}
\end{figure}

Let us turn to the SS junction in the  "topological" parameter range.
Solving Eq. \ref{eq:spectrum}  gives the spectrum of Andreev states (Fig. \ref{fig5}b). We observe the level crossing at $E=0,\phi=\pi$ being lifted in a narrow energy interval. Strikingly, we observe another pair of levels with energies remaining small in the whole range of phase. These levels are absent in TNTT picture and emerge as a consequence of topological triviality of the s-matrix. Since there is no level crossing at $E=0$, the parity of the ground state is always even. 

The situation can be comprehended if we notice that each matrix $\hat{s}_1,\hat{s}_2$
forming the resulting $\hat{s}$ brings a resonant pole corresponding to a "buried" zero-energy state at far end of each wire. The $\hat{s}$ thus has two resonant poles. The mixing of the two "buried" states  results in their (phase-dependent) energy splitting and formation of the pair of low-energy Andreev levels. The eigenvalues of s-matrix
move in the narrow energy interval reconnecting next-to-nearest (two poles) neighbour "high-energy" eigenvalues (Fig.\ref{fig4}b). This brings four rather than two states in the vicinity of the
crossing point $E=0,\phi=\pi,\chi=0$, all being involved in the lifting of the degeneracy.
 The detailed theory of the crossing point will be presented elsewhere.

Since the s-matrix remains topologically trivial,  there can be no sharp transition in its characteristics  that would correspond to the "topological" transition in the (rather unphysical) limit of infinite wire. However, a BdG s-matrix with resonant poles  is characterized by a topological number that can change sharply upon changing the parameters.

Let us illustrate this with a two-pole scattering matrix  correspond to the fork setup in Fig.
5 a. Here the scattering matrices $S_1$, $S_2$ of fork tines bring a resonant pole
each.  The BdG symmetry leaves two distinct  possibilities for  the poles of the total scattering
matrix :$i$.  both poles lie on the imaginary energy axis ($E=-i\Gamma_1,\,-i\Gamma_2$), $ii$. they
form a pair symmetric with respect to reflection ${\rm Re} E \to -{\rm Re} E$
($E=\pm\varepsilon-i\Gamma$). One can now change the s-matrix $\check{S}_0$ describing the normal
scattering in the fork. If the tines are open to the lead states, the pole configuration should be like one for two parallel SN junctions: the possibility $ii$ is realized. If the tines are isolated, the "buried"
states mix resulting in an energy spitting: the possibility $i$ is realized. We thus expect and
\cite{supplementary} prove  the transition at intermediate coupling.

Generally, one can characterize a BdG s-matrix of arbitrary dimension with a topological number that
is just the number of poles lying precisely on the imaginary axis. We expect this number to change
by $2$ upon changing the parameters, this gives a series of "topological" transitions. (Fig. 5 b,c) Two poles
are degenerate at the transition point. However, since in general the degenerate poles are at
finite imaginary energy $\Gamma$, the manifestations of the transitions in transport properties are
limited. The energy-dependent Andreev conductance does not seem to have a singularity at the
transition point.


We have performed the topological analysis of the properties of SN and SS junctions characterized by BdG s-matrices. We have proven topological triviality of physical matrices that describe finite-size junctions: there is neither TNT, nor TNTT. This implies the absence of a sharp "topological"  transition upon crossing to "topological" parameter range as well as the absence of $4\pi$-periodic Josephson effect. We have resolved the apparent contradiction with results of \cite{Lutchyn,Oreg,Fu2} by considering the low-energy poles of s-matrices. The resulting sharp energy dependence at $E\approx 0$leads  to  Lorentian energy dependence of Andreev conductance. We have demonstrated a topological transition (or a series of transitions) of a different kind associated with a change of the configuration of the resonant poles in complex energy plane.

This research was supported by the Dutch Science Foundation NWO/FOM. The authors are indebted to 
C. W. J. Beenakker, C. Kane, R. M. Lutchyn, F. von Oppen and L. P. Kouwenhoven for useful discussions.

\section{Appendix}
We consider the fork setup presented in Fig. 5 a of
the main text.
Let us specify the normal-scattering matrix $\tilde{S}_0
$ that determines the reflection of electrons and holes coming from the lead and their transmission to the fork tines. In general form, we can write this matrix in blocks  of reflection and transmission matrices:
\begin{equation}
\tilde{S}_0 = \left(
\begin{array}{cc}
\check{R}_d & \check{T}_{ud}\\
\check{T}_{du} & \check{R}_u
\end{array}
\right).
\end{equation}
with $\check{R}_{d}$($\check{R}_{u}$) being 
the reflection matrix to the lead (to the tines).
We denote the scattering matrix of the two tines as $\check S={\rm diag} \left\{ \hat S_1, \hat S_2 \right\}$. The total scattering
matrix then reads:
\begin{eqnarray}
\check S_{\rm tot} = \check R_d + \check T_{du} \check S \check T_{ud} + \check T_{du} \check S \check R_{u}
\check S \check T_{ud} +
\ldots =\nonumber\\
=\check R_d + \check T_{du} \check S \frac{1}{\check 1 - \check R_{u} \check S}\check T_{ud}.
\end{eqnarray}
The poles of $S_{\rm tot}$ are determined by the zeros of the determinant in the above expression,
\begin{equation}
\det \left(\check 1 - \check R_u \check S \right)=0. \label{eq:poles}
\end{equation}
Now we implement the pole decomposition (Eq. 2 of the main text) for each $\hat S_i$. Using the fact that  $\check R_u$ is invertible we obtain:
\begin{eqnarray}
\det \left(\check R_u^{-1}\check S_p^{-1}  - {\rm diag}\left[ \left(\hat 1 -
\frac{2i\Gamma_1} {E +
i\Gamma_1} |\psi_{1}><\psi_{1}|\right),\nonumber\right. \right.\\
\left.\left. \left(\hat 1 - \frac{2i\Gamma_2} {E + i\Gamma_2} |\psi_{2}><\psi_{2}|\right)\right]\right)=0,
\end{eqnarray}
\begin{figure}
\centerline{\includegraphics[width=0.9\linewidth]{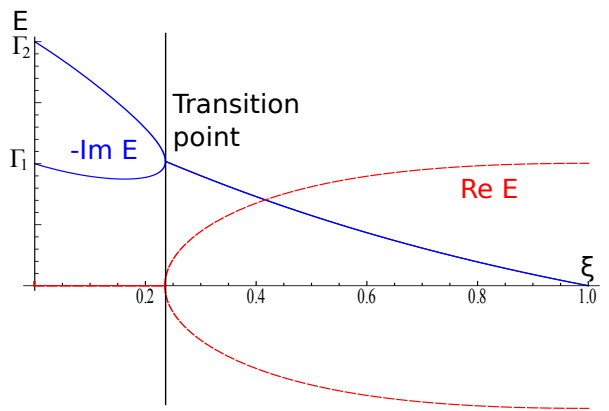}} \caption{Pole positions versus parameter $\xi$ characterizing the model setup. At $\xi=0$ there is no reflection in the fork, while at $\xi=1$ the reflection is complete so that the tines are isolated. We chose $\Gamma_2=2 \Gamma_1$.}
 \label{fig6}
\end{figure}
\begin{figure}
\centerline{\includegraphics[width=0.9\linewidth]{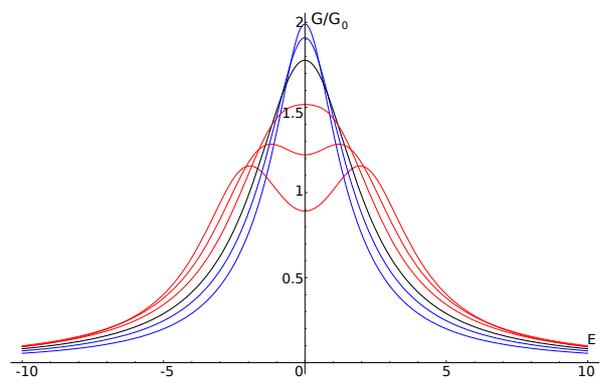}} \caption{Andreev conductance of the fork setup versus energy for a set of parameter values(from the top curve at $E=0$ to the bottom one): $\xi=0.05,0.15,0.24$ (corresponds to the transition, thick line) $0.35,0.45,0.55$, $G_0=2e^2/h$.} \label{fig7}
\end{figure}
with $S_p$ being the energy-independent (at $E\simeq\Gamma$) part of the s-matrix of the tines. We use convenient notations $E\equiv i\epsilon$ and invert the matrix $\check R_u^{-1}\check S_p^{-1} - \check 1$ to arrive at
\begin{eqnarray}
\det\left(\check 1 + \frac{\check S_p \check R_u}{\check 1 - \check S_p \check R_u}\times \nonumber\right. \\ \times \left.{\rm diag}\left[ \frac{2\Gamma_1} {\epsilon + \Gamma_1}|\psi_1><\psi_1| , \frac{2\Gamma_2} {\epsilon + \Gamma_2}|\psi_2><\psi_2| \right]\right)=0.
\end{eqnarray} 
The condition of zero determinant can be then expressed in terms of the matrix elements in the basis 
$|\psi_2>,|\psi_1>$ that we arrange into a $2 \times 2$ matrix
 $\tilde M$:
\begin{eqnarray}
\tilde M = \left<\begin{array}{c}
\psi_1\\
\psi_2
\end{array}\right| \frac{\check S_p \check R_u}{\check 1 - \check S_p \check R_u} \left|\begin{array}{c}
\psi_1\\
\psi_2
\end{array}\right>.
\end{eqnarray}
and reads
\begin{eqnarray}
\det\left(
\begin{array}{cc}
1 + M_{11} \frac{2\Gamma_1} {\epsilon + \Gamma_1} & M_{12}\frac{2\Gamma_1} {\epsilon + \Gamma_1}\\
M_{21} \frac{2\Gamma_2} {\epsilon + \Gamma_2} & 1 + M_{22} \frac{2\Gamma_2} {\epsilon + \Gamma_2}
\end{array}
\right)=0. \label{eq:M} 
\end{eqnarray}
This reduces to a quadratic equation
\begin{eqnarray}
\epsilon^2 + \epsilon(\Gamma_1 + 2\Gamma_1 M_{11} + \Gamma_2 + 2\Gamma_2 M_{22}) + \nonumber \\
+\Gamma_1 \Gamma_2 ((1+2M_{11})(1+2M_{22}) - 4 M_{12} M_{21})=0.
\end{eqnarray}
We note that all matrix elements are real owing to the BdG symmetry. This is why the equation roots are either real or mutually conjugated. The intermediate situation between these two cases is two degenerate real roots.
This is the point of topological transition. 

Let us prove that the topological transition  of this kind
should inevitably occur upon changing the $S_0$
from full transmission to full reflection.

 In case of full transmission. $\check R_u\to \check 0$. Therefore, $\check{M} \to 0$ and  the roots $\epsilon\to \Gamma_1$ or $\to \Gamma_2$. Therefore, we are in the situation with two real roots. 
The case of full transmission is a bit more difficult to handle. In this case, $R_u$ is a unitary matrix. With this, 
$$
\frac{\check S_p \check R_u}{\check 1 - \check S_p \check R_u} = -\frac{\check{1}}{2} +\frac{i}{2} \cot(\check{h})
$$
$\check{h}$ being a Hermitian matrix. The BdG symmetry requires $\check{h}$ to be asymmetric in Majorana basis. Since the pole eigenvectors
$|\psi_1\rangle,|\psi_2\rangle$ are real in this basis,
$M_{11}=M_{22} = -1/2$ and $M_{12} = - M_{21}$. With this,
the quadratic equation reduces to 
\begin{equation}
\epsilon^2 + 4 \Gamma_1 \Gamma_2 |M_{12}|^2 =0.
\end{equation}
that has two conjugated (and purely imaginary) roots.

 This implies that changing from full transmission to full reflection  requires passing a point when two roots are degenerate, so that, the point of the topological transition. 

We  illustrate with a simple  model involving $4 \times 4$ s-matrices. We choose 
\begin{eqnarray}
\check R_u = -\check R_d = \frac{\xi}{\sqrt{2}} \left(\begin{array}{cc} 1 & 1\\ -1 & 1 \end{array}\right);\\
\check T_{ud} = \check T_{du} = \frac{\sqrt{1-\xi^2}}{\xi} \check R_u.
\end{eqnarray}
Since these matrices describe normal scattering, they are diagonal in electron-hole space. The parameter value $\xi =0$($\xi=1$) corresponds to the case of full transmission (reflection) We choose the scattering matrix of the tines  to be 
\begin{eqnarray}
S_p=\left(
\begin{array}{cccc}
\frac{E}{E+i\Gamma_1} & 0 & \frac{i\Gamma_1}{E+i\Gamma_1} & 0 \\
0 & \frac{E}{E+i\Gamma_2} & 0 & \frac{i\Gamma_2}{E+i\Gamma_2} \\
\frac{i\Gamma_1}{E+i\Gamma_1} & 0 & \frac{E}{E+i\Gamma_1} & 0 \\
0 & \frac{i\Gamma_2}{E+i\Gamma_2} & 0 & \frac{E}{E+i\Gamma_2}
\end{array}
\right),
\end{eqnarray}
Here, two upper rows are for electron part of the wave while two lower rows are for hole part. This is the simplest matrix with two poles at $-i\Gamma_1, -i\Gamma_2$. In Fig. \ref{fig6} we plot the positions of poles versus $\xi$.  The topological transition takes place at $\xi \approx 0.24$.

We plot  the energy dependence of Andreev conductance for this model setup at several values of $\xi$ in Fig. \ref{fig7}.
Qualitatively, one expects a single-peak energy dependence in the case of
big transmission, and a double-peak dependence in the case of low transmission, the positions of the peaks corresponding to the energy of Andreev bound state. This is indeed seen in the Figure. A naive expectation would be that the intermediate situation where the second derivative of Andreev conductance vanishes, occurs at the topological transition. Yet this does not happen.

It remains unclear at the moment whether the topological transition under consideration manifests itself as a singularity of any physical quantity. While this is likely the case in the model of non-interacting electrons, the interactions may change this.

\begin{thebibliography}{100}

\bibitem{Josephson}
B. D. Josephson, Rev. Mod. Phys. {\bf 36}, 216 (1964);
A. F. Andreev, Zh. Teor. Eksp. Fiz. \textbf{46}, 1823 (1964) [Sov. Phys. JETP \textbf{19},
1228 (1964)].


\bibitem{Nazarov1}
Y. V. Nazarov and Y. M. Blanter, \textit{Quantum Transport}, (Cambridge University Press, Cambridge, 2009).

\bibitem{Shelankov}
A. L. Shelankov, Zh. Eksp. Teor. Fiz., Pis'ma. \textbf{32}, 2, 122-125 (1980); G. E. Blonder, M. Tinkham, and T. M. Klapwijk, Phys. Rev. B \textbf{25}, 4515 (1982).

\bibitem{Beenakker2}
C. W. J. Beenakker, Phys. Rev. Lett. \textbf{67}, 3836 (1991).

\bibitem{de Jong}
M. J. M. de Jong and C. W. J. Beenakker, Phys. Rev. Lett. {\bf 74}, 1657 (1995);
N. M. Chtchelkatchev \textit{et al.}, 2001, Pis’ma Zh. Eksp. Teor. Fiz. \textbf{74}, 357–361 (2001) [JETP Lett. \textbf{74}, 323–327 (2001)];
O. V. Dimitrova and M. V. Feigel’man, J. Exp. Theor. Phys. \textbf{102}, 652 (2006);
C. Bruder, Phys. Rev. B \textbf{41}, 4017 (1990);
Y. Tanaka and S. Kashiwaya, Phys. Rev. Lett. \textbf{74}, 3451 (1995).


\bibitem{Kitaev}
A. Y. Kitaev,   Phys.-Usp. {\bf 44}, 131–136 (2001),  arXiv:cond-mat/0010440 (unpublished).

\bibitem{Sau}
J. D. Sau \textit{et al.}, Phys. Rev. Lett. \textbf{104}, 040502 (2010).

\bibitem{Lutchyn}
R. M. Lutchyn, J. D. Sau, S. Das Sarma, Phys. Rev. Lett. \textbf{105}, 077001 (2010).

\bibitem{Oreg1}
Y. Oreg, G. Refael, and F. von Oppen,  Phys. Rev. Lett. {\bf 105}, 177002 (2010). 

\bibitem{Duckheim}
L. Fu and C.L. Kane, Phys. Rev. Lett. \textbf{100}, 096407 (2008);
M. Duckheim, P. W. Brouwer, arXiv:1011.5839 (2010),unpublished ; 
S. B. Chung \textit{et al.}, arXiv:1011.6422 (2010), unpublished.

\bibitem{Beenakker3}
C. W. J. Beenakker, Rev. Mod. Phys. \textbf{69}, 731 (1997).

\bibitem{Ryu}
S. Ryu et al., New J. Phys. \textbf{12}, 065010 (2010).

\bibitem{Fu2}
L. Fu and C. L. Kane, Phys. Rev. B \textbf{79}, 161408(R) (2009).

\bibitem{Oreg}
J. Alicea \textit{et al.}, Nature Physics doi:10.1038/nphys1915 (2011).

\bibitem{Merz}
F. Merz and J. T. Chalker, Phys. Rev. B \textbf{65}, 054425 (2002);
A. R. Akhmerov \textit{et al.}, Phys.Rev.Lett. {\bf 106}, 057001 (2011) 

\bibitem{Hassler}
F. Hassler \textit{et al.}, New J. Phys. \textbf{12}, 125002 (2010).

\bibitem{Beenakker4}
C. W. J. Beenakker \textit{et al.}, Phys.Rev.B {\bf 83}, 085413 (2011).

\bibitem{Wigner}
E. Wigner, Phys. Rev. \textbf{70}, 15 (1946).

\bibitem{Majorana}
E. Majorana, Nuovo Cimento \textbf{14}, 170 (1937).

\bibitem{Falko}
S.V.Kuplevakhskii and I. I. Falko, Pis’ma Zh. Eksp.
Teor. Fiz. {\bf 52}, 957 (1990) [JETP Lett. {\bf 52}, 340 (1990)].

\bibitem{supplementary}
See appendix for details of the proof.
\end{thebibliography}
\end{document}